\let\oldequation\equation
\let\oldendequation\endequation
\renewcommand{\equation}{\oldequation*}
\renewcommand{\endequation}{\oldendequation*}

\documentclass[12pt]{iopart}

\expandafter\let\csname equation*\endcsname=\relax
\expandafter\let\csname endequation*\endcsname=\relax
\usepackage{amsmath}
\usepackage{subcaption}
\usepackage{iopams}
\usepackage{booktabs}
\usepackage{lineno}
\usepackage{graphicx} 
\usepackage{dcolumn} 
\usepackage{bm} 
\usepackage{hyperref} 
\usepackage{color}
\usepackage{multirow}
\usepackage{footnote}
\usepackage{listings}
\usepackage{adjustbox}
\usepackage{caption}

\begin{document}

\title[]{Rapid Parameter Estimation for Merging Massive Black Hole Binaries Using Continuous Normalizing Flows}

\author{Bo Liang$^1 \, ^2 \, ^3 \, ^4$, Minghui Du*$^1$, He Wang*$^4 \, ^6$, Yuxiang Xu$^1 \, ^2 \, ^3 \, ^4$, Chang Liu$^7$, Xiaotong Wei$^1$, Peng Xu$^1 \, ^2  \, ^4 \, ^5$, Li-e Qiang$^7$, Ziren Luo$^1 \, ^2 \, ^4 \, ^6$}
\address{1 Center for Gravitational Wave Experiment, National Microgravity Laboratory, Institute of Mechanics, Chinese Academy of Sciences, Beijing 100190, China}
\address{2 Key Laboratory of Gravitational Wave Precision Measurement of Zhejiang Province, Hangzhou Institute for Advanced Study, UCAS, Hangzhou 310024, China}
\address{3 Shanghai Institute of Optics and Fine Mechanics, Chinese Academy of Sciences, Shanghai 201800, China}
\address{4 Taiji Laboratory for Gravitational Wave Universe (Beijing/Hangzhou), University of Chinese Academy of Sciences (UCAS), Beijing 100049, China}
\address{5 Lanzhou Center of Theoretical Physics, Lanzhou University, Lanzhou 730000, China}
\address{6 International Centre for Theoretical Physics Asia-Pacific (ICTP-AP), University of Chinese Academy of Sciences (UCAS), Beijing 100049, China}
\address{7 National Space Science Center, Chinese Academy of Sciences, Beijing 100190, China}

\ead{hewang@ucas.ac.cn, duminghui@imech.ac.cn}
\vspace{10pt}

\begin{abstract}
Detecting the coalescences of massive black hole binaries (MBHBs) is one of the primary targets for space-based gravitational wave observatories such as LISA, Taiji, and Tianqin. 
The fast and accurate parameter estimation of merging MBHBs is of great significance for the global fitting of all resolvable sources, as well as the astrophysical interpretation of gravitational wave signals.
However, such analyses usually  entail significant computational costs.
To address these challenges, inspired by the latest progress in generative models, we explore the application of continuous normalizing flows (CNFs) on the parameter estimation of MBHBs. 
Specifically, we employ linear interpolation and trig interpolation methods to construct transport paths for training CNFs. 
Additionally, we creatively introduce a parameter transformation method based on the symmetry in the detector's response function. This transformation is integrated within CNFs, allowing us to train the model using a simplified dataset, and then perform parameter estimation on more general data, hence also acting as a crucial factor in improving the training speed.
In conclusion, for the first time, within a comprehensive and reasonable parameter range, we have achieved a complete and unbiased 11-dimensional rapid inference for MBHBs in the presence of astrophysical confusion noise using CNFs. In the experiments based on simulated data, our model produces posterior distributions comparable to those obtained by nested sampling.


\end{abstract}

\section{Introduction}

Gravitational wave (GW) astronomy has made significant progress due to the breakthroughs in ground-based detections led by the LIGO-Virgo-KAGRA network~\cite{aasi2015advanced,acernese2014advanced,kagra2019kagra}. Scheduled in the upcoming decade, space-based detectors such as the Laser Interferometer Space Antenna (LISA)~\cite{amaro2017laser,baker2019laser}, Taiji~\cite{hu2017taiji,luo2021taiji,2021CmPhy...4...34T_simplified}, and Tianqin~\cite{luo2016tianqin, Luo_2016} aim to detect GWs in the 0.1 mHz - 1 Hz frequency band associated with enormous astrophysical and cosmological sources~\cite{heinzel2003interferometry}.
As one of the representative projects, 
Taiji consists of a triangle of three spacecraft (S/Cs) with a baseline separation of 3 million kilometers and aims to detect low-frequency GWs emitted by sources such as compact galactic binaries (GBs)~\cite{zhang2022resolving}, massive black hole binaries (MBHBs)~\cite{klein2016science}, extreme mass ratio inspirals (EMRIs)~\cite{maselli2020detecting}, as well as the stochastic gravitational wave background of astrophysical or cosmological origins.

{\color{black}The merger of MBHB with component masses $10^4\sim 10^7 M_{\odot}$ is one of the primary observation targets of space-based GW detectors, since they may shed light upon the growth and merger history of massive black holes, the dynamic behavior of curved spacetime, and the nature of gravity, etc.
Due to the presence of the surrounding gas environment, MBHBs may produce detectable electromagnetic (EM) signals in addition to GWs before, during, and after the mergers~\cite{EM1,EM2,EM3,EM4}. 
Joint EM and GW observation of the same source may unveil the mystery of the evolution of MBHBs and their host galaxies, and provide a new measurement of the cosmic expansion rate, hence shedding light upon the Hubble tension problem in cosmology~\cite{Hubble,HubbleTension}. 
For these purposes, it is crucial that the estimation of MBHB parameters, especially the sky location (ecliptic longitude, latitude), should be conducted with high precision and low latency, to guide the search of EM observatories.}

Besides, the parameter estimation of merging MBHB must also be robust against the contamination of the confusion foreground formed by tens of millions of Galactic binaries. 
Indeed, in the area of space-based GW detection, data processing faces the challenge caused by the presence of a large number of astrophysical sources in the measurement band~\cite{bayle2022overview}.
The non-trivial overlap of individual signals in the time and frequency domains increases the complexity of the task.
The accurate analysis of up to $\mathcal{O}(10^4)$ overlapping sources in the data requires the use of global fit techniques~\cite{cornish2005lisa,littenberg2020global,littenberg2023prototype}, namely jointly fitting all sources and noise in the data, which imposes a high challenge on both computational time and resources.
The prototype global fit pipeline proposed by Ref.~\cite{littenberg2023prototype} currently requires as many as $\mathcal{O}(10^3)$ CPUs and several days to process the 12-month Lisa Data Challenge (LDC) data, and the issue of computational resources would become even more critical for the multi-year datasets in the future.  
Due to these challenges and problems, the aim of this paper is to develop an accelerated and robust method to estimate the parameters of merging MBHBs.

When discussing the efficiency and speed of parameter estimation, the advancement in machine learning technology is revolutionizing our understanding of this area. Currently, most machine learning methods applied to the merging of MBHB use neural posterior estimation (NPE)~\cite{Papamakarios2016FastI,Lueckmann2017FlexibleSI,Greenberg2019AutomaticPT} of discrete normalized flows (DNFs). NPE has been widely used in fields such as atmospheric retrievals (AR)~\cite{Gebhard2023InferringAP} and ground-based gravitational wave parameters estimation~\cite{Dax2021RealtimeGS,Green2020GravitationalwavePE}. NPE was also used by the winning entry to the 2023 edition of the ARIEL data challenge~\cite{Aubin2023SimulationbasedIF}.

The mission lifetimes of LISA, Taiji, and Tianqin are planned to last for years, during which the payloads, orbits, and noise characteristics will vary over time. Accordingly, the models used for parameter estimation may also need to be regularly updated. As a result, the training speed of the model will be linked to the timeliness of scientific output.
However, a significant challenge to the current NPE model in MBHB inference is the extensive training duration, which hampers its ability to conduct real-time data analysis. 
Additionally, 
previous research~\cite{Du2023AdvancingSG} in this area shows that NPE falls short in accurately reproducing the posterior distribution obtained by the benchmark Nested Sampling method. 
This prolonged training period not only delays the application of the model but also limits its scientific utility in rapidly evolving research contexts.

To address these challenges, we first apply Flow Matching (FM)~\cite{lipman2022flow} with Continuous Normalizing Flows (CNFs) to MBHBs. 
FM, as a training objective for CNFs, offers more flexibility in designing non-diffusion paths, such as optimal transport paths, and enables direct access to density functions~\cite{Dax2023FlowMF}. This method is called Flow Matching Posterior Estimation (FMPE).
The basic implementation of FM~\cite{lipman2022flow} uses  simple linear interpolants to construct optimal transport paths.
In our work, We also utilize trig interpolants~\cite{Albergo2023StochasticIA, tong2024improving} to design transport paths that allow for the continuous evolution of MBHB waveform features along the time axis for modeling. We call this method Variance Preserving Flow Matching Posterior Estimation (VPFMPE).

\begin{figure}[h]
  \centering
  \includegraphics[width=1.0\textwidth]{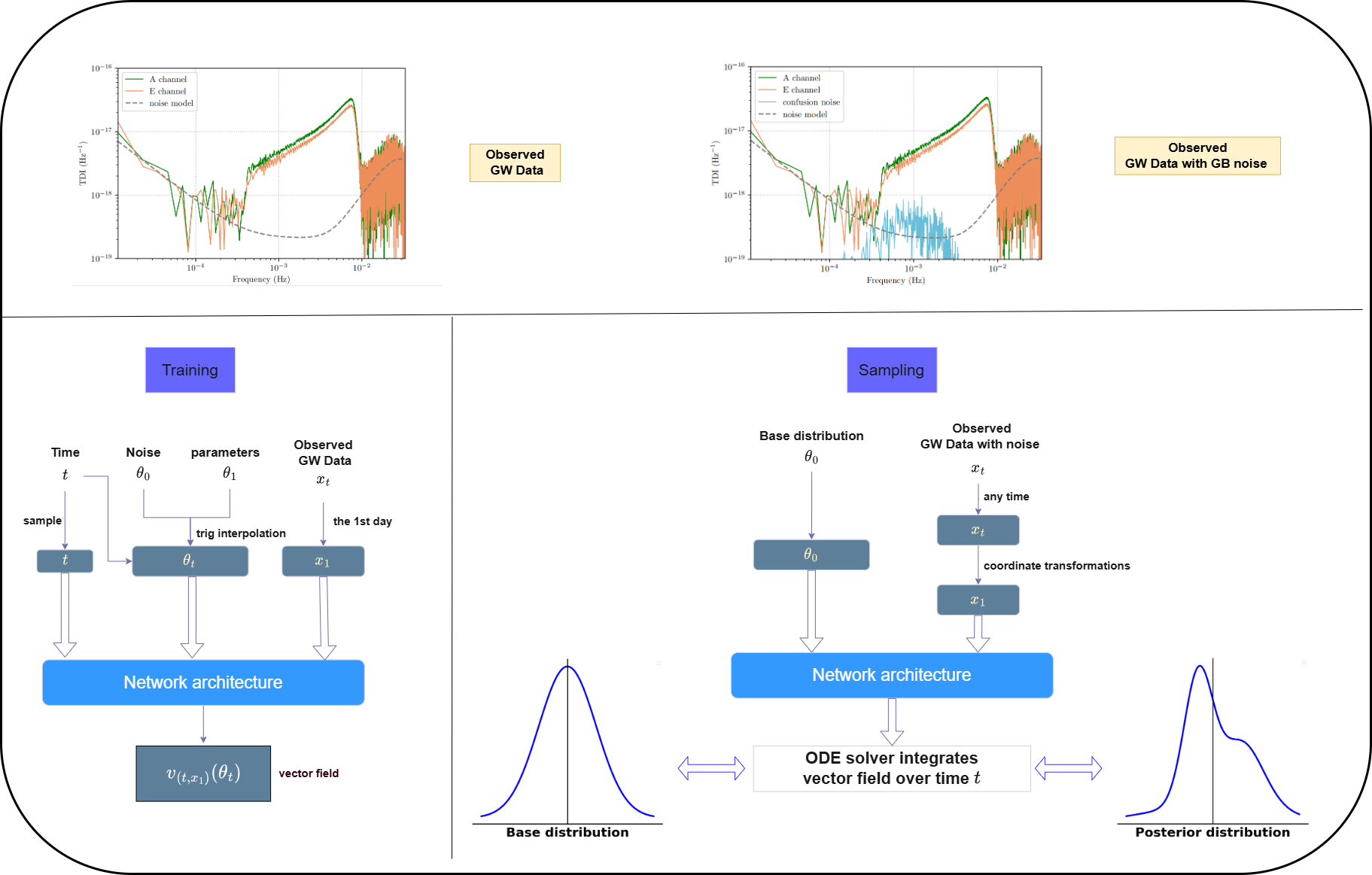}
  \captionsetup{font={footnotesize,stretch=0.4}}
  \caption{
   The Diagrams for the training and sampling phases of our model. On the lower left, the training phase is displayed, where $t$ denotes time and $x_1$ represents the data from the first day of observed MBHBs. In this phase, a feature extractor, as detailed in section~\ref{Network}, processes the data to produce a final vector field. This vector field enables the network architecture to learn a transformation from a base distribution ($\theta_0$, sampled from a Gaussian distribution) to a posterior distribution ($\theta_1$, sampled from the posterior distribution). On the lower right, the sampling phase is shown. Here, the data with noise ($x_t$), which includes both confusion noise and instrumental noise from MBHB observations at any time, is mapped back to the first day's data $x_1$ using the coordinate transformation described in section~\ref{Model Generalization Strategy}. The Gaussian distribution is then converted into a posterior distribution by solving the ordinary differential equation (ODE) with the vector field learned during the training phase. 
   The top of the image displays simulation data from both the training and sampling phases. 
  }
  \label{fig:model}
\end{figure}

The challenges of parameter estimation for MBHBs not only involve the need for sophisticated algorithms capable of feature extraction and inference, but also lie in the excessively wide prior range of parameters and the resulting complexity of data. Recent studies~\cite{RUAN2023137904, Du2023AdvancingSG} adopted priors that do not fully cover the complete physical ranges of parameters. This compromise was due to that setting broad prior ranges during model training can reduce the effectiveness of the model. This underscores the importance of finding a careful balance between the breadth of prior ranges and the accuracy of estimation. To date, parameter estimation based on the full prior range of MBHB still remains to be explored. 
While, our method is proven to overcome this limitation and achieve  unbiased parameter  estimation on the full prior range, hence improving the applicability of AI models. 

In order to further enhance the efficiency and effectiveness of model training, we tackled a  complexity commonly faced by space-based gravitational wave data analysis, namely the time-varying response function. Utilizing the coordinate transformation of parameters, we can essentially conduct both the model training and the parameter estimation in a coordinate frame fixed to the detector, in which the response function is time-invariant. This strategy significantly simplifies the training task. 
Consequently, the integration of coordinate transformation and advanced generative models enhances the speed and accuracy of parameter estimation for MBHBs. 

At last but not least, the network architecture developed in this work, named TResMlp, for GW feature extractions is illustrated in Fig~\ref{fig:model_res}, which introduces the innovation of encoding time $t$, resulting in superior training speed and accuracy compared to existing methods~\cite{Dax2023FlowMF, Du2023AdvancingSG, RUAN2023137904}. The training of the model can be finished in about two days, as shown in Figure~\ref{fig:model}, which is roughly three times faster than the training duration in previous work~\cite{Du2023AdvancingSG}.

In conclusion, we performed a complete unbiased 11-dimensional fast inference of  MBHB mergers using FM with CNFs. Compared to traditional Monte Carlo methods, we are at least several orders of magnitude faster, while still producing similar posterior distributions. 

\section{Data Generation}\label{sec:DataSimulation}

In this section, we introduce the simulation of datasets used for model training and testing. 
This part of  work is a development  based on our previous methods~\cite{Du2023AdvancingSG}, with the most noteworthy improvement being  the  extension of  prior ranges.

For the waveform template, we choose the IMRPhenomD template which describes the dominant $(\ell, |m|) = (2, 2)$ harmonic  of inspiral, merger and ringdown waveform for non-precessing binary black holes~\cite{PhenomD1,PhenomD2}. 
To suppress laser frequency noise which dominates the raw interferometric data, a pre-processing technique known as Time Delay Interferometry (TDI) was proposed~\cite{CancellationTinto1999,TimeTinto2002}, therefore we calculate Taiji's response to MBHB signals in the form of two noise-orthogonal TDI channels $(A, E)$~\cite{ThePrince2002}.
 The fast Fourier domain response function derived by Ref.~\cite{Marsat:2018oam} is employed to project the IMRPhenomD waveform to the $(A, E)$ TDI channels. The TDI response function can be written in a concise form as  
\begin{equation}\label{eq:FrequencyDomainResponse}
    \tilde{A}(f) = \sum_\alpha \mathcal{T}^{A}_\alpha(f, t_f) \tilde{h}_\alpha(f), 
    \quad      
    \tilde{E}(f) = \sum_\alpha \mathcal{T}^{E}_\alpha(f, t_f) \tilde{h}_\alpha(f),  
\end{equation}
where $\alpha$ denotes the polarization mode of waveform ($\alpha \in \{+, \times\}$).  
$\mathcal{T}_\alpha^{A,E}$ depend on the orbital motion of detector via the time-frequency relationship $t_f$  
\begin{equation}
    t_f = -\frac{1}{2\pi}\frac{{\rm d} \Psi(f)}{{\rm d} f},
\end{equation}
with $\Psi(f)$ being the Fourier domain phase of the IMRPhenomD waveform. 
The motion of detector is  modeled based on an equal-arm analytic orbit with nominal arm-length of $3\times 10^9 \ {\rm m}$ and orbital period of 1 year. The Frequency domain TDI responses are calculated using a modified version of the open-source code \texttt{BBHx}~\cite{Katz:2020hku,Katz:2021uax}.


For the model considered in this paper, the parameter space of  MBHB is 11-dimensional, whose symbols as well as descriptions can be found in Table~\ref{table:priors}. Also listed in Table~\ref{table:priors} are the prior ranges of parameters. 
The explanations for the settings of   priors are as follows. 
Similar to the LDC-sangria data, the prior range of $\mathcal{M}_c$ covers an order of magnitude. 
In our previous work~\cite{Du2023AdvancingSG}, to reduce the complexity of data, we limited the prior ranges of spins to $\chi_{z1, 2}\in [0, 1]$, and 
NPE was demonstrated to achieve unbiased estimations on this simplified parameter estimation task.
However, the performance of NPE became  subpar when applied to the  complete spin priors (i.e. $\chi_{z1,2} \in [-1, 1]$). 
One of the key advantages of the models proposed in this paper is that they are applicable to the complete spin priors.
Since the merger stage of each MBHB appears as a prominent peak in the time domain, we set the prior range of $t_c$ to be within $\sim 2000$ s ($\approx$ 0.02 day), with the central value denoted as the ``reference time'' $t_{\rm ref}$. 
Notably, to simplify the dataset and reduce the difficulty of training, in the training set, 
$t_{\rm ref}$ is fixed to the $1^{\rm th}$ day. The response function varies with the motion of detector with a 1-year period, thus a training set with $t_c \in [1-0.01, 1+0.01]$ day is far from being sufficient to represent all the observable MBHBs. 
While, thanks to the method described in Sec.~\ref{Model Generalization Strategy}, our model can be generalized to make inferences on MBHB mergers throughout the entire mission lifetime. 
Consequently, when generating the test set, $t_{\rm ref}$ can take any value within the mission time. 
For other extrinsic parameters, the range of $d_L$ corresponds to the redshift range of $z \in [1, 10]$, 
and the angles $\iota$ and $\beta$ are uniformly distributed in terms of $\cos\iota$ and $\sin \beta$, since the sources are isotropically oriented and located.

\begin{table}[htb]
    \centering
    \footnotesize
    \small
    \captionsetup{font={footnotesize,stretch=0.4}}
    \label{table:prior}
    \caption{The prior ranges of parameters.     
    The table shows the range of prior distributions for each parameter. 
    }\label{table:priors}
    \begin{adjustbox}{width=\linewidth}
    \begin{tabular}{ccccc}
    \toprule
        \textbf{Parameter}  & \textbf{Description}  &  \textbf{Prior Lower Bound} & \textbf{Prior Upper Bound} \\
        \hline
        $\mathcal{M}_c$  & Chirp mass of the binary system &$1.0 \times 10^5 M_\odot$ & $1.0 \times 10^6 M_\odot$ \\
        $q$  & Ratio of masses of the two black holes & 0.1 & 1.0 \\
        $\chi_{z1}$   & Spin of the 1st black hole along the $z$-axis & -1 & 1 \\
        $\chi_{z2}$ & Spin of the 2nd black hole along the $z$-axis & -1 & 1  \\
        $t_c$ & Time of coalescence relative to the reference time & -0.01 day & 0.01 day \\
        $\varphi_c$ & Phase at the moment of coalescence & 0 rad & $2\pi$ rad \\
        $d_L$ & Luminosity distance to the binary system & 6000 Mpc & 100000 Mpc \\
        $\iota$ & Angle of inclination of the binary orbit & 0 rad & $\pi$ rad \\
        $\beta$ & Ecliptic latitude of the binary system & $-\frac{\pi}{2}$ rad & $\frac{\pi}{2}$ rad \\
        $\lambda$ & Ecliptic longitude of the binary system & 0 rad & $2\pi$ rad \\
        $\psi$ & Polarization angle of the gravitational wave & 0 rad & $\pi$ rad \\
        \bottomrule
    \end{tabular}
    \end{adjustbox}
\end{table}

The majority of a MBHB's signal-to-noise ratio (SNR) is contributed by a short duration containing the merger stage (typically several hours)~\cite{pratten2023lisa}, 
therefore 
the time duration for each sample is set to one day.
Besides, we set the sampling frequency of data to $1/15$ Hz, and   we have checked that a Nyquist frequency of 0.033 Hz is beyond the maximum frequencies of all waveforms within the priors.

\begin{figure}[htb]
  \centering
  \includegraphics[width=0.8\textwidth]{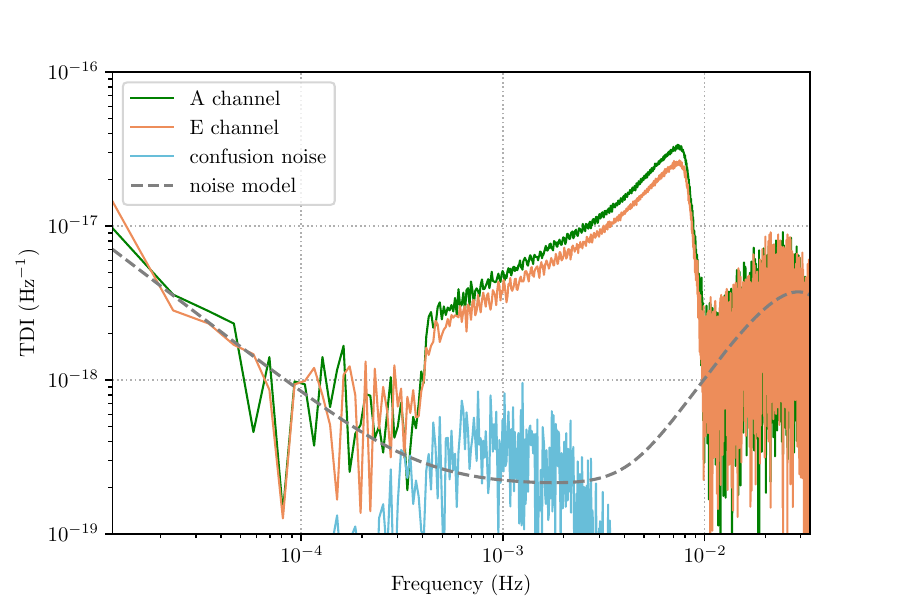}
  \captionsetup{font={footnotesize,stretch=0.4}}
  \caption{A representative sample in the test set. The orange and green curves represent the Fourier transforms of the TDI-A, E data streams, which include GW signals, instrumental noises, and confusion noise. To clearly illustrate the relative strengths of the signal and noises, the blue curve depicts the Fourier transform of the confusion noise, while the gray dashed curve indicates the theoretical amplitude of the instrumental noises.
  }
  \label{fig:noise}
\end{figure}

Each sample in the training set is the combination of signal and  instrumental noises. 
According to the current requirement of Taiji, 
{\color{black}the one-sided amplitude spectral densities (ASDs) of the two main  instrumental noises, namely optical metrology system noise and test-mass acceleration noise, are }~\cite{luo2020brief}
{\begin{eqnarray}\label{eq:TaijiNoise}
    \sqrt{S_{\mathrm{OMS}}(f)}= & 8 \times 10^{-12}  \sqrt{1+\left(\frac{2 \mathrm{mHz}}{f}\right)^4} \frac{{\mathrm{m}}}{\sqrt{\mathrm{Hz}}}, \\
    \sqrt{S_{\mathrm{ACC}}(f)}= & 3 \times 10^{-15}   \sqrt{1+\left(\frac{0.4 \mathrm{mHz}}{f}\right)^2} \sqrt{1+\left(\frac{f}{8 \mathrm{mHz}}\right)^4} \frac{{\mathrm{m/s^2}}}{\sqrt{\mathrm{Hz}}},
\end{eqnarray}
and the ASDs of total instrumental noises of the $A, E$ channels (dubbed 
 $\sqrt{S_n^{A, E}(f)}$) are the combinations of $S_{\mathrm{OMS}}(f)$ and $S_{\mathrm{ACC}}(f)$ (see Ref.~\cite{Vallisneri:2007xa,Wang:2020vkg,Wang:2020fwa} for their specific forms). 
We not only incorporated simulated noise into the MBHB signals during the training process but also utilized Taiji's ASD function as additional contextual information fed to the model.

As is mentioned before, one of the advantages of our model is its robustness to the confusion noise. 
In the  test set, we further include the  foreground originating from the unresolvable GW signals of $\sim 3\times10^7$  GBs. 
The simulation of confusion noise follows the methods of Refs.~\cite{PhysRevD.76.083006,Wang:2023jct,2019MNRAS.483.5518K,liu2023confusion,Zhang:2022wcp}.

Shown in Figure~\ref{fig:noise} is a representative sample in the test set including GW signal, instrumental noises, and confusion noise.

\section{Methodological Framework}

The application of machine learning techniques in the field of gravitational wave data analysis has seen widespread success and remarkable achievements~\cite{jordan2015machine, Cuoco_2021}. Among these techniques, deep learning has demonstrated exceptional prowess, especially in processing the data of ground-based detectors~\cite{gabbard2022bayesian,chatterjee2019using,green2020gravitational,green2021complete,delaunoy2020lightning,krastev2021detection,shen2021statistically,dax2021real,dax2023neural}. 
Next, we will introduce the theoretical concepts of discrete and continuous flows, as well as the corresponding training objectives.

\textbf{NPE with DNFs}.
NPE combined with DNFs is used to directly fit the posterior distribution \( p(\theta | x) \). In this approach, an invertible mapping \( \psi_x \) transforms the base distribution \( p_0(\mathbf{z}) \) into the target distribution \( q(\theta | x) \):

\begin{equation}
q(\theta | x) = p_0(\psi_x^{-1}(\theta))   \det \left|\left( \frac{\partial \psi_x^{-1}(\theta)}{\partial \theta} \right) \right|.
\end{equation}

NPE is trained using a maximum likelihood objective and simplifies the expectation using Bayes' theorem.
\begin{equation}
L_{\mathrm{NPE}} = - E_{p(\theta)p(x | \theta)} \log q(\theta | x).
\end{equation}

Once training is complete, NPE can efficiently perform inference on new observations \( x \) and provide exact density evaluations of \( q(\theta | x) \). This method retains the expressive power and computational efficiency of normalizing flows while achieving posterior fitting.

\textbf{CNFs for parameter estimation}.
CNFs transform a simple base distribution into a more complex distribution by describing this transformation through continuous time \( t \in [0, 1] \). For each \( t \), the flow is defined by a vector field \( v_{t,x} \) on the sample space. This vector field represents the trajectory velocity of the samples at any given time \( t \).
\begin{equation}
    \frac{d}{dt} \psi_{t,x}(\theta) = v_{t,x}(\psi_{t,x}(\theta)).
\end{equation}
The conversion between the base distribution (at \( t = 0 \)) and the target distribution (at \( t = 1 \)) is achieved by integration. A straightforward approach to training CNFs involves starting with a given base distribution and using an ODE solver to obtain the target distribution. This target distribution is then constrained to match the real data distribution by minimizing a divergence measure, such as the KL divergence. 
Because many intermediate trajectories are unknown, inferring the distribution (either through sampling or likelihood computation) requires repeatedly simulating the ODE, leading to a significant computational burden.

However, Flow matching provides a training objective for CNFs, Dax et al.~\cite{Dax2023FlowMF} apply flow matching~\cite{lipman2022flow} to simulation-based inference. The loss function of FMPE is expressed as:
\begin{equation}
L_{FMPE} = E_{t \sim p(t), \theta_1 \sim p(\theta), x \sim p(x|\theta_1), \theta_t \sim p_t(\theta_t|\theta_1)} \parallel v_{t,x}(\theta_t) - u_t(\theta_t|\theta_1) \parallel^2. \ \ 
\end{equation}

Here, $v_{t,x}(\theta_t)$ represents the vector field that generates the path to the target probability distribution, and $u_t(\theta_t|\theta_1)$ is the sample-conditioned vector field. FMPE introduces a Gaussian path with time-dependent means \( t\theta_{1} \) and standard deviations $1 - (1 - \sigma)t $. 
Overall, $t$ is sampled from $p(t)$ ($t \in [0, 1]$), and the training data is generated by sampling $\theta_1$ from the prior and then simulating the corresponding data $x$ for $\theta_1$~\cite{Dax2023FlowMF}. The sample-conditional probability path must satisfy that at $t=0$, it is the base distribution (i.e., near $\theta_0$), and at $t=1$, it is the target distribution (i.e., near $\theta_1$).

\begin{equation}
p_{t}(\theta|\theta_{1}) = N(t\theta_{1},1 - (1 - \sigma)t ). 
\end{equation}
FMPE defines the sample-conditioned vector field is 
\begin{equation}
u_t(\theta|\theta_1) = \frac{\theta_1 - (1-\sigma)\theta_{0}}{1 - (1 - \sigma)t}. 
\end{equation}

In our work, based on the studies by Albergo et al.~\cite{Albergo2023StochasticIA,tong2024improving} and utilize triangular interpolation to define the time-dependent mean and standard deviation, thus constructing the Gaussian path and sample-conditioned vector field as follows:
\begin{equation}
p_{t}(\theta|\theta_{1}) = N(\sin(\frac{\pi t}{2})\theta_{1} + \cos(\frac{\pi t}{2})\theta_{0},\sigma). 
\end{equation}

\begin{equation}
u_t(\theta|\theta_1) =\frac{\pi}{2(\cos(\frac{\pi t}{2})\theta_{1} - \sin(\frac{\pi t}{2})\theta_{0})}.
\end{equation}

We also use it to train continuous normalization flows. We show the differences between the two methods in Table ~\ref{table:vp}.

 \begin{table}[h]
    \scriptsize
    \centering
    \captionsetup{font={footnotesize,stretch=0.4}}
    \caption{A table with constructing transport paths differences. FMPE uses linear interpolation to construct transport paths, while VPFMPE uses trig interpolants to construct transport paths.}
    \label{table:vp}
    \begin{tabular}{ccccccc}
    \hline
    \multicolumn{1}{l}{} &
      \multicolumn{1}{l}{Interpolar Method} &
      \multicolumn{1}{l}{Mean of the probability path} &
      \multicolumn{1}{l}{Standard deviation of the probability path} &
       \\ \hline
    FMPE &
      linear &
      $t\theta_{1}$ &
      $1 - (1 - \sigma)t$ &
       \\
    VPFMPE &
      trig &
      $\sin(\frac{\pi t}{2})\theta_{1} + \cos(\frac{\pi t}{2})\theta_{0}$ &
      $\sigma$ &
       \\ 
       \hline
    \end{tabular}
    \label{tab:mytable}  
\end{table}

\subsection{Network architecture}\label{Network}

A key challenge in space-based gravitational wave parameter estimation lies in managing the high dimensionality  of  data. To address this, it is essential to develop a feature extraction network capable of transforming the high-dimensional input data into a more manageable, low-dimensional representation~\cite{Dax2021RealtimeGS,Du2023AdvancingSG,dingo}. This transformation is critical for adjusting the dimensionality of the final feature vector to match that of the parameter vector $\theta$, enabling the network to effectively parameterize the conditional vector field $(t,x,\theta) \rightarrow v_{t,x}(\theta)$~\cite{Dax2023FlowMF}. Achieving accurate parameterization of this conditional vector field is fundamental to the construction of an effective network structure.
\begin{figure}[htb]
  \centering
  \includegraphics[width=1.0\textwidth]{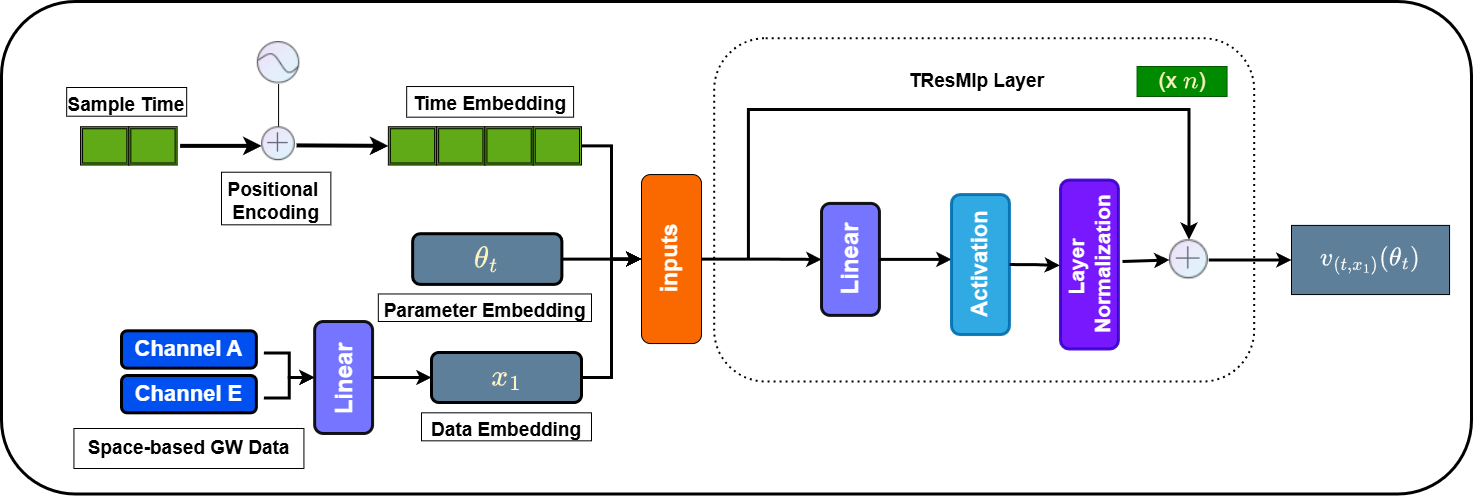}
  \captionsetup{font={footnotesize,stretch=0.4}}
  \caption{
  TResMlp Network Architecture: Initially, the frequency vector $f$ is used to encode the position of time $t$. Sine and cosine functions are then applied to extract periodic features from time $t$, resulting in the ``Time Embedding". Subsequently, a ``Linear" layer maps the two channels of the space-based GW data to produce the ``Data Embedding". Finally, all obtained embeddings are input into a Multi-Layer Perceptron (MLP) with a residual structure to generate the final output \( v_{(t,x_1)}(\theta_t) \).}
  \label{fig:model_res}
\end{figure}
In recent years, DeResNet networks have been commonly used as the backbone network for feature extraction in ground-based or space-based gravitational wave parameter estimations~\cite{dax2021real, Du2023AdvancingSG,Dax2023FlowMF, green2021complete}, and Gated Linear Unit(GLU)-conditioning~\cite{2016arXiv161208083D} can be optionally used for information aggregation of $(\theta,t)$.

However, if information about the actual parameter $\theta$ is not properly handled or retained during the process of aggregating $(\theta,t)$ information using GLU, it could potentially lead to a loss of information about $\theta$. 
To encode the time $t$ more effectively, we introduce a frequency vector $f$. This frequency vector $f$ adapts dynamically to changes in the parameters of the wave source, allowing for a more precise encoding of time $t$. Our network architecture, referred to as TResMlp, is shown in Figure~\ref{fig:model_res}.


We trained the CNFs of MBHBs using two different network architectures. To ensure fairness, the two models had identical parameters such as the optimizer and learning rate, differing only in their network structures. Additionally, both architectures used the same hidden layer configuration. We show in Figure~\ref{fig:resmlp} a comparison of the posterior estimates of the four parameters of the injected MBHBs signal by the two models.


\begin{figure}[htb]
  \centering
  \includegraphics[width=0.8\textwidth]{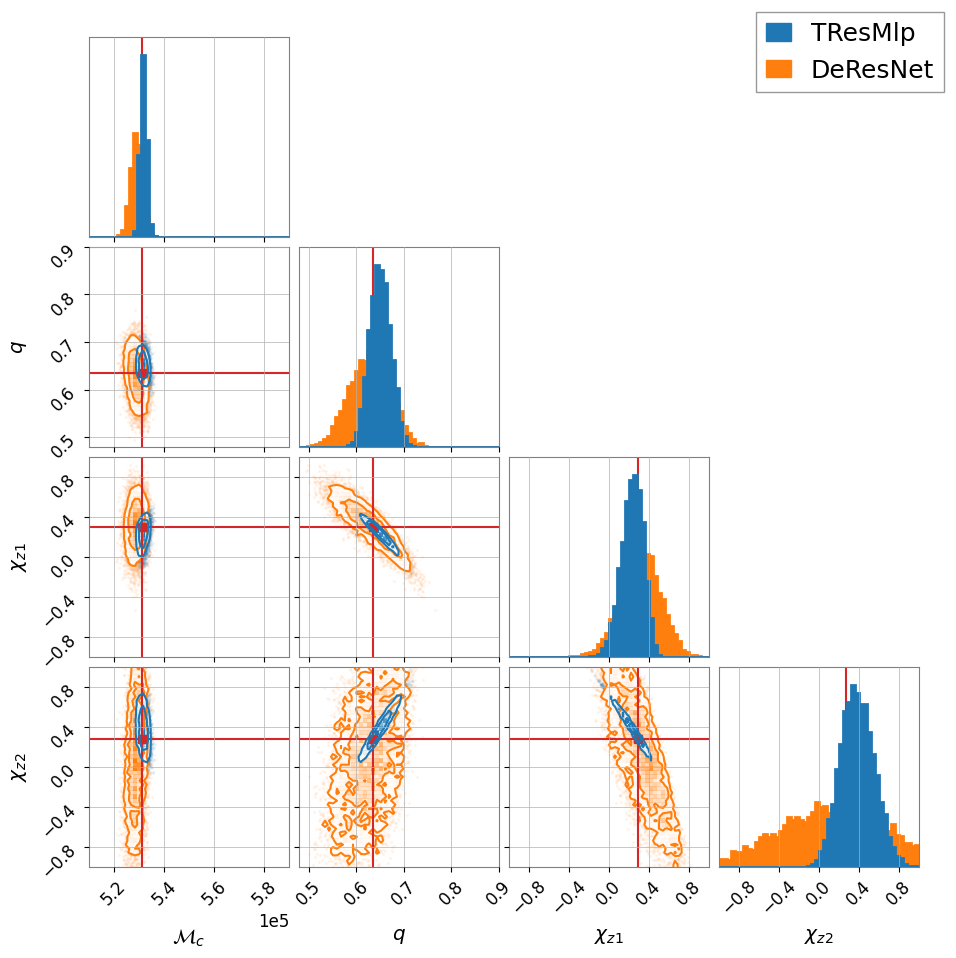}
  \captionsetup{font={footnotesize,stretch=0.4}}
  \caption{Comparison of the posterior distributions of the two network architectures. The blue represents the posterior distribution plot of CNFs trained using the TResMlp network. The orange represents the posterior distribution plot of CNFs trained using the DeResNet network. The red line represents the true value of the MBHBs.
  }
  \label{fig:resmlp}
\end{figure}
Compared to the DeResNet model that uses GLU-conditioning for aggregating information about $(\theta, t)$~\cite{dax2021real, Du2023AdvancingSG, Dax2023FlowMF, green2021complete}, our network architecture, TResMlp, demonstrates faster training speed and higher precision with the same hidden layers. More experiments and comparisons of related models are described in detail in Section~\ref{Experiments}.

\subsection{Coordinate Transformation and Model Generalization Strategy}\label{Model Generalization Strategy}  

To simplify the dataset and reduce the difficulty of training, in the training set, the coalescence times $t_c$ of MBHBs are set within a small interval $[1-0.01 , 1+0.01]$ day, 
see Sec.~\ref{sec:DataSimulation}. 
In principle, the model trained on this dataset can only be applied to MBHBs that merge within this range.
While, using the generalization strategy introduced 
 in this section, the application scope of the trained model can be  extended to the whole mission lifetime. 

The complexity of space-based GW data partly lies in the annual variation of response function due to the orbital motion of detector constellation around the sun. 
In other words, the mapping between data and parameters varies with $t_f$, see Eq.~(\ref{eq:FrequencyDomainResponse}).  Since the time duration of data in consideration is much smaller than the time scale of orbital motion, 
$t_f$ can be well approximated by $t_{\rm ref}$, 
thus $t_{\rm ref}$ labels the position of detector in the orbit. 
The time-dependence of response function significantly increases the complexity of the training set, and  severely damage the training efficiency and the performance of the model. 
To tackle this challenge, we exploit the symmetry (invariance) of the system, fix $t_{\rm ref}$ in the training data, and then the trained model is generalized via coordinate transformation to make inferences at any time. 
\begin{figure}[htb]
  \centering
  \includegraphics[width=1.0\textwidth]{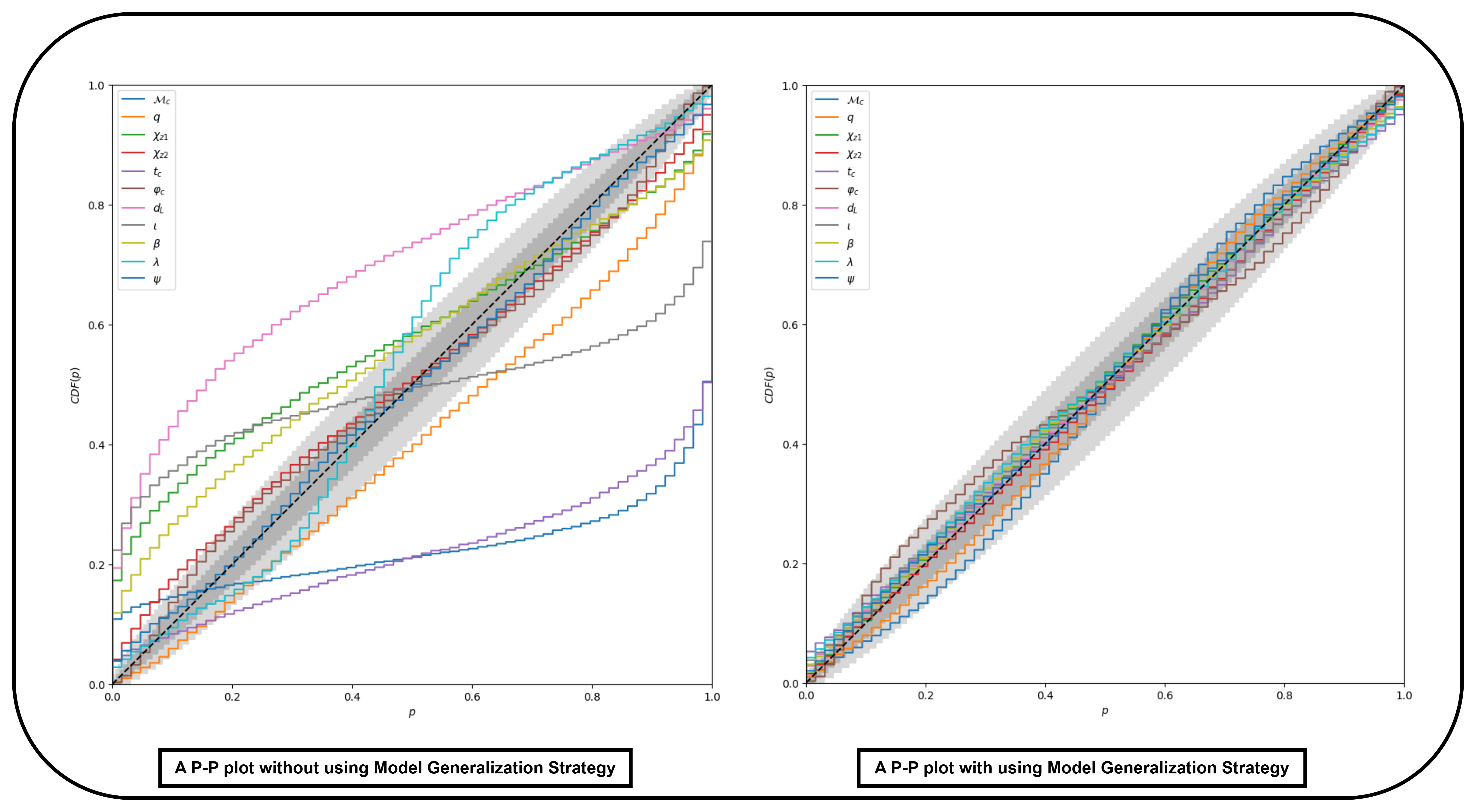}
  \captionsetup{font={footnotesize,stretch=0.4}}
  \caption{Comparison of the effect of the model generalization strategy on model performance. The left panel shows the P-P plot without the model generalization strategy, while the right panel shows the P-P plot with the model generalization strategy. Both plots are for a set of 4000 injections with added instrument noise and reference times ranging from 1 to 365 days. Ideally, the cumulative distribution function of these percentiles for each parameter should align closely with the diagonal line, indicating optimal network performance. In such a case, the percentiles would be uniformly distributed between 0 and 1. The grey regions on the plot represent the  $1\sigma$ and $2\sigma$ confidence intervals.
  }
  \label{fig:p-p plot}
\end{figure}

To explain the so called ``symmetry'', we define two coordinate systems, namely the Solar System Barycenter (SSB) frame, in which parameters are indexed with subscript ``S'', and a coordinate frame fixed to the Taiji detector, in which parameters are indexed with subscript ``T''. 
Parameters which differ in these two  coordinate systems are $\{\lambda, \beta, \psi\}$.
As is explained, in the SSB frame, the mapping between parameters $\bm{\theta}_S$ and TDI response varies with $t_{\rm ref}$. While in the Taiji frame, due to the short duration of data, the motion of GW source relative to the detector can be neglected, thus the mapping between parameters $\bm{\theta}_T$ and TDI response is independent of $t_{\rm ref}$.  Essentially, the former $t_{\rm ref}$-dependent mapping can be regarded as the combination of the latter $t_{\rm ref}$-independent mapping and a $t_{\rm ref}$-dependent coordinate transformation of parameters between the SSB frame and Taiji frame, and the analytical formula of this coordinate transformation is known, given the information of orbit. 
This is the fundamental principle behind training data simplification and model generalization. 

Following Refs.~\cite{OptimalKrolak2004,CoordinateTransformation}, we denote as $O_1(\lambda, \beta, \psi)$ the rotation matrix between the polarization basis in the GW source frame and the observer frame (Taiji or SSB), and $O_2(t_{\rm ref})$ the rotation matrix between the SSB frame and Taiji frame (see Eq.~(6) and Eq.~(9) of Ref.~\cite{OptimalKrolak2004} for their expressions). Parameters $\{\lambda, \beta, \psi\}$ can be transformed between Taiji and SSB frames by solving~\cite{CoordinateTransformation}:
\begin{align}\label{eq:13}
    [\cos \lambda_T \cos \beta_T, \sin \lambda_T \cos \beta_T, \sin \beta_T] &= O_2^T(t_{\rm ref}) [\cos \lambda_S \cos \beta_S, \sin \lambda_S \cos \beta_S, \sin \beta_S], \nonumber \\ 
    O_1(\lambda_T, \beta_T, \psi_T) &= O_2^T(t_{\rm ref}) O_1(\lambda_S, \beta_S, \psi_S).
\end{align}
For brevity, we further denote the $t_{\rm ref}$-dependent coordinate transformation from Taiji frame to SSB frame as $\mathcal{F}_{t_{\rm ref}}: \bm{\theta}_T\rightarrow\bm{\theta}_{S, {t_{\rm ref}}}$. 

The strategy of model generalization is described as follows: During the training stage, the model is  trained on a dataset with $t_c \in [t_{\rm ref0}-0.01, t_{\rm ref0}+0.01]$, where $t_{\rm ref0} \equiv 1$. When using the trained model to perform parameter estimation on the test data (generally $t_{\rm ref} \neq t_{\rm ref0}$),  we firstly obtain the estimated parameters $\bm{\theta}_{S, t_{\rm ref 0}}$. Obviously these parameters are not the desired ones. 
Then, the desired parameters $\bm{\theta}_{S, t_{\rm ref}}$ can be  deduced following two steps:
\begin{equation}
    \bm{\theta}_{T} = \mathcal{F}_{t_{\rm ref 0}}^{-1} \left(\bm{\theta}_{S, t_{\rm ref 0}} \right), \quad
    \bm{\theta}_{S, t_{\rm ref}} = \mathcal{F}_{t_{\rm ref }} \left(\bm{\theta}_{T} \right).
\end{equation}
By performing these steps on each posterior sample, we finally obtain the target 
 posterior distribution
\begin{equation}\label{eq:mapping}
    p\left(\bm{\theta}_{S, t_{\rm ref}}|d\right) = p\left(\bm{\theta}_{S, t_{\rm ref 0}}|d\right) J\left(\mathcal{F}^{\ }_{t_{\rm ref 0}}\right)J\left(\mathcal{F}^{-1}_{t_{\rm ref}}\right),
\end{equation}
with $J\left(\mathcal{F}^{\ }_{t_{\rm ref 0}}\right)$ and $J\left(\mathcal{F}^{-1}_{t_{\rm ref}}\right)$ being the Jacobian determinants corresponding to $\mathcal{F}_{t_{\rm ref 0}}$ and $\mathcal{F}^{-1}_{t_{\rm ref}}$, respectively, and $d$ stands for the data. Formula~(\ref{eq:mapping}) looks similar to the transformations in normalizing flows, except that mapping $\mathcal{F}_{t_{\rm ref}}$ is constructed analytically, instead of training a neural network. 
As a result, our knowledge about the system has greatly simplified the training task.  
This idea represents a paradigm of combining physical information with neural networks.
Methods belonging to this family, such as the group-equivariant neural posterior estimation~\cite{gnpe,dingo}, have shown bright prospects.
}
For future space-based missions, once the detectors’ motions are obtained through orbit determination, we can calculate the specific forms of matrices $O_1$ and $O_2$, thereby applying Eq.~(\ref{eq:13}) - Eq.~(\ref{eq:mapping}) to conduct the model generalization strategy.

Below, we describe how models can be combined with the model generalization strategy. We employ VPFMPE as the loss function for training CNFs. We utilize the network architecture TResMlp as the gravitational wave feature extractor, labeling this approach as VPFMPE-TResMlp. In the training phase we only use the 1st day's data. During the inference phase, we employ the model generalization strategy to obtain the desired posterior distribution. This allows the model to be trained using the  simplified  dataset  but tested and applied to data in the whole mission lifetime. We use the P-P plot in Figure~\ref{fig:p-p plot} to demonstrate the importance of using a model generalization strategy.

\subsection{Comparison of methods}
In this section, we injected a substantial number of MBHB signals with noise to a comparison between the efficacy of FM with CNFs and NPE with DNFs (Recent advances in machine learning for parameter estimation in MBHBs). We use the three mentioned methods for FM with CNFs to compare them with NPE with DNFs. we employed the four models mentioned above: VPFMPE-TResMlp, FMPE-TResMlp, FMPE-DeResNet, and NPE-DeResNet~\cite{Du2023AdvancingSG}. We applied the model generalization strategy~\ref{Model Generalization Strategy} to all four models by default. Otherwise, due to the complexity of the dataset, the models would lack inference capability.

To ensure the fairness of the training process, all models were trained under uniform conditions, involving 400 epochs, with each batch containing 14,336 samples.
All models were trained using a single NVIDIA A800 GPU.
In terms of optimization strategy, we used a combination of cosine annealing~\cite{loshchilov2016sgdr} and the Adam optimiser~\cite{kingma2014adam} to gradually reduce the learning rate to zero over the course of the training process.
Our model training is based on a dataset consisting of 5 million MBHB signals. To prevent overfitting, we reserved 5\% of the dataset for validation purposes. During the training process, no signs of overfitting were detected. 
During the testing phase of the model, we performed random sampling based on the prior distributions defined in Table~\ref{table:priors} and generated MBHB signals for testing.

\begin{table}[t]
\scriptsize
\captionsetup{font={footnotesize,stretch=0.4}}
\caption{In the 850 simulations, we compared the correlation coefficients between the true values and the recovered values for the four models. In addition, in the last two columns of the table, we list the network parameters and training times for the three models.}
\label{table:vs}
\begin{tabular}{ccccccccc}
\hline
\multicolumn{2}{l}{}            & $\mathcal{M}_c$ & $q$    & $\chi_{z1}$ & $\chi_{z2}$ & $d_L$  & Network parameters  &   Training time  \\ \hline
\multirow{3}{*}{CNFs} & VPFMPE-TResMlp & 0.9978          & 0.9790 & 0.9723      & 0.8163      & 0.9408 & $\approx$ 189 millions & $\approx$ 2.0 days \\
    & FMPE-TResMlp  & 0.9978 & 0.9787 & 0.9719 & 0.8139 & 0.9400 & $\approx$ 189 millions & $\approx$ 2.0 days \\
    & FMPE-DeResNet & 0.9970 & 0.9762 & 0.9563 & 0.7174 & 0.9397 & $\approx$ 211 millions & $\approx$ 2.3 days \\ \hline
DNFs & NPE-DeResNet  & 0.9261 & 0.8647 & 0.6620 & 0.1681 & 0.8634 & $\approx$ 384 millions & $\approx$ 6.5 days \\ \hline
\end{tabular}
\end{table}
We injected 850 waveforms drawn from the prior shown in Table~\ref{table:priors}, along with residual instrumental and confusion noise. 
We will take the median of the recovered values as the point estimate and calculate the correlation coefficient with the true values of the parameters. The correlation coefficient reflects the degree of linear correlation between the true value distribution and the estimated values. A coefficient value of 1 indicates perfect agreement between the estimated and true values.
Note that the reference times of these testing signals are randomly distributed throughout a year, rather than being fixed to the 1st day.

Our results are presented in Table~\ref{table:vs}. In the final column of the Table~\ref{table:vs}, we have included the approximate training times for the four models. For $t_c$, $\varphi_c$, $\iota$, $\beta$, $\lambda$, and $\psi$, due to the presence of multimodality (i.e. there is more than one peak in the posterior distributions), using the median as a point estimate is not applicable.
Our results indicate that FM with CNFs reduced training time by a factor of three and achieved complete unbiased estimation for the injected signals. In contrast, NPE with DNF not only had a longer training time but also produced significant distortions in $\chi_{z1}$ and $\chi_{z2}$, potentially leading to erroneous scientific analysis. Among the methods we tested CNF for MBHBs, the VPFMPE-TResMlp model achieved the best results and exhibited the fastest training time.

\section{Experiments} \label{Experiments}

Based on the analysis of P-P plot (Figure~\ref{fig:p-p plot}), we confirmed that the model can achieve unbiased estimates when  considering only the instrumental noises.
However, the confusion noise, which is composed of the GW signals from tens of millions of GBs, is also one of the main noise sources in the mHz frequency band. Currently, an accurate GB population model is still absent, and the characterization of confusion noise is typically conducted after the identification of bright signals such as MBHBs in the whole global fitting pipeline. Considering these practical factors, we have not included confusion noise in the training set. Instead, in order to assess the model’s applicability under real detection scenarios, a simulated confusion noise is incorporated in the test set. 

To date, existing studies have not been successful in using likelihood-free inference to provide unbiased estimates of the MBHB across the entire parameter space.
For example, Ref.~\cite{RUAN2023137904} only focused on four parameters, and Ref.~\cite{Du2023AdvancingSG} have limited to positive spins.
Also, we use FM with CNFs for complete parameter estimation of MBHBs and achieve comparable results to nested sampling.
Therefore, in subsequent comparisons, we will only compare the results of VPFMPE-TResMlp to those of the MCMC method.
\begin{figure}[htb]
  \centering
  \includegraphics[width=0.8\textwidth]{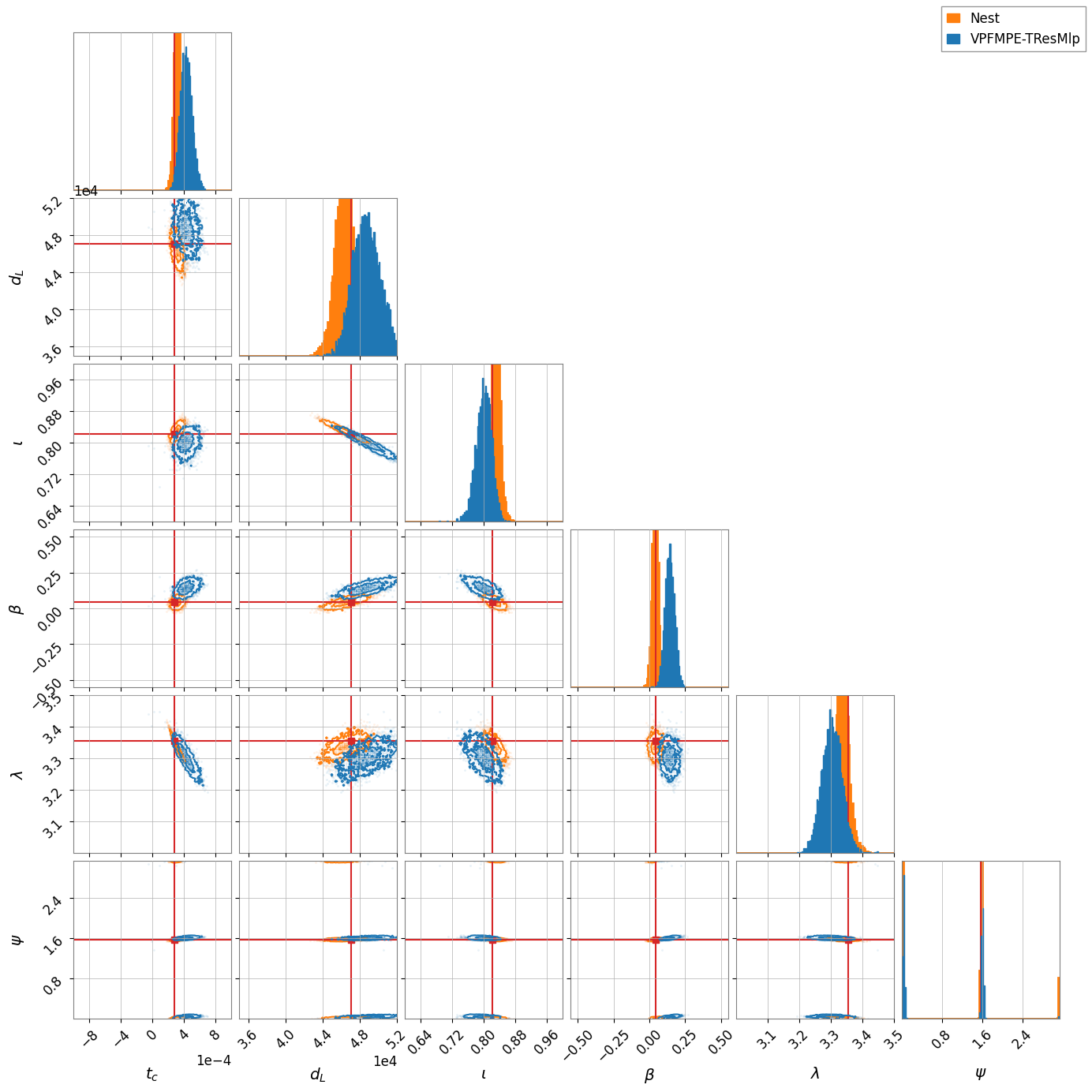}
  \captionsetup{font={footnotesize,stretch=0.4}}
  \caption{
Corner plot of parameter posterior distributions for gravitational wave signals injected with instrumental and confusion noise. The blue contours illustrate the two-dimensional joint posteriors obtained from our (VPFMPE-TResMlp) model, with peaks selected based on $t_c$ that align with true values. In contrast, the orange contours represent the corresponding posteriors obtained from benchmark analysis using Nested Sampling. For each model, the contour boundaries respectively encompass 95\% and 60\% confidence levels. 
The true parameter values of the simulated signal are indicated by red vertical and horizontal lines.
  }
  \label{fig:corner_extrinsic}
\end{figure}
The aim of using CNFs for MBHBs is to reproduce the comparable posterior distribution as the MCMC method. Thus, by comparing the results of Bayesian inference on the same event, the performance of the model can be better assessed.
In this case, we use Nested Sampling~\cite{veitch2010bayesian,skilling2006nested,ashton2019bilby} as a Monte Carlo technique for comparison experiments, as shown in Figure~\ref{fig:corner_extrinsic}.

An example using the same prior probabilities outlined in Table~\ref{table:priors} is provided in Figure~\ref{fig:corner_extrinsic}, where a test injection is subjected to confusion noise on the 30th reference day. The injected GW signal originates from an MBHB with a total SNR of 1543, and its parameters are detailed in Table~\ref{table:injection}.
Our model exhibited results similar to the traditional nested sampling, demonstrating consistency in the sense of the degeneracy among extrinsic parameters. This consistency aligns with the crucial physical characteristics of MBHBs. Note that compared to nested sampling, our machine learning process always produces similar posterior distributions on $t_c$, $d_L$, $\iota$, $\lambda$, and $\psi$, as shown in Figure~\ref{fig:corner_extrinsic}.
We have achieved accurate estimates of the sky position parameters. This is most advantageous since accurate and reliable estimates of the sky position parameters (longitude, latitude) are useful in guiding the search for electromagnetic observatories.

\begin{table}[htb]  
    \captionsetup{font={footnotesize,stretch=0.4}}
    \caption{This table compares the parameters injected in the prior with the parameters recovered by the Nested Sampling and VPFMPE-TResMlp methods. The GW signal is emitted by a merging MBHB with a total SNR of 1543. The recovered values are accompanied by their 1 $\sigma$ confidence regions.}
    \small
    \label{table:injection}
    \centering
    \begin{tabular}{cccc}
        \toprule
        \textbf{Parameter} & \textbf{Injected value} & \textbf{VPFMPE-TResMlp} & \textbf{Nest} \\
        \hline
        $\mathcal{M}_c$ [$M_\odot$]   & 531067.4  &  $528444.8000_{-1206.2360}^{+1141.3900}$ & $531038.3974_{-107.2206}^{+125.0852}$  \\
        $q$   & 0.6355  & $0.5994_{-0.0220}^{+0.0236}$  & $0.6405_{-0.0070}^{+0.0060}$ \\
        $\chi_{z1}$ & 0.2930 &     $0.4051_{-0.1042}^{+0.0858}$  & $0.2636_{-0.0259}^{+0.0318}$  \\
        $\chi_{z2}$ & 0.2726 & $0.0273_{-0.1552}^{+0.1807}$ & $0.3233_{-0.0517}^{+0.0409}$  \\
        $t_c$ [day]  & 0.00028 & $0.0004_{-0.0001}^{+0.0001}$ & $0.0003_{-0.0000}^{+0.0000}$ \\
        $\varphi_c$ [rad] & 0.5412 & $3.1765_{-1.8449}^{+1.7528}$ & $3.6714_{-3.0978}^{+1.6062}$ \\
        $d_L$ [Mpc] & 47054.1 & $48698.8670_{-1486.6328}^{+1568.7870}$ & $46334.8753_{-891.2652}^{+882.4137}$ \\
        $\iota$ [rad] & 0.8207 & $0.8004_{-0.0207}^{+0.0191}$ & $0.8298_{-0.0109}^{+0.0098}$ \\
        $\beta$ [rad] & 0.0459 & $0.1405_{-0.0317}^{+0.0340}$ & $0.0401_{-0.0188}^{+0.0188}$ \\
        $\lambda$ [rad] & 3.3544 & $3.3024_{-0.0322}^{+0.0317}$ & $3.3350_{-0.0167}^{+0.0188}$ \\
        $\psi$ [rad] & 1.5634 & $1.5680_{-1.5388}^{+0.0442}$ & $1.5692_{-1.5613}^{+0.0188}$ \\
        \bottomrule
    \end{tabular}
\end{table}

Specifically, under the influence of confusion noise, we have collated the true values of the injections alongside the results generated by both the model and the nested sampling process in Table ~\ref{table:injection}.
The test results show that, except for the parameters $\beta$, $\chi_{z1}$ and $\chi_{z2}$, our machine learning model is consistent with the posterior distributions of nested sampling in terms of order of magnitude. 
However, we also observed a decrease in the model's performance in estimating $\beta$, $\chi_{z1}$ and $\chi_{z2}$ under the influence of confusion noise, but they are generally still within the 2 $\sigma$ or 3 $\sigma$ range.
Such accuracy of estimation is basically sufficient for a rapid preliminary parameter estimation, and manifests the robustness of our model when applied to data outside the training set.
Addressing the issue of confusion noise will be regarded as a key focus of our future work, in order to further enhance the robustness and accuracy of the model.

In the final part of the results, we explore the computational efficiency of the algorithm. Table~\ref{table:vss} shows the speed of generating posterior samples using nested sampling compared to the VPFMPE method. It is important to note that for nested sampling the search is performed near the near the true intrinsic parameter values of the injected MBHB. In practice, the time consumed by nested sampling is significantly longer than the times listed in the table. Even under these conditions, our VPFMPE method remains two orders of magnitude faster than the nested sampling method. 
Additionally, the VPFMPE method employed an NVIDIA A800 GPU for sampling, while nested sampling utilized an Intel(R) Xeon(R) Platinum 8268 CPU.

 \begin{table}[h]
    \scriptsize
    \centering
    \captionsetup{font={footnotesize,stretch=0.4}}
    \caption{Comparison of computational time required by different posterior sampling approaches to generate their respective samples.}
    \label{table:vss}
    \begin{tabular}{ccccccc}
    \hline
    \multicolumn{1}{l}{} &
      \multicolumn{1}{l}{Number of Posterior Samples} &
      \multicolumn{1}{l}{Runtime (seconds)} &
      \multicolumn{1}{l}{Time per Sample} &
       \\ \hline
    Nested Sampling &
      11215 &
      3000 &
      0.26750 &
       \\
    VPFMPE &
      14336 &
      240 &
      0.01674 &
       \\ 
       \hline
    \end{tabular}
\end{table}

Our model not only prioritizes efficiency but also generates a posterior distribution similar to nested sampling. Moreover, it is well-suited for the majority of MBHB sources anticipated by future Taiji observations. This contribution will significantly enhance global fit analysis.

\section{Conclusions}

In this study, we have demonstrated the inaugural application of using CNFs for MBHBs in the field of space-based GW detection, taking the parameter estimation of merging MBHBs as a representative example. We recommend the FM with CNFs, particularly the VPFMPE-TResMlp method, as a promising approach for the inference of merging MBHBs.

It is shown that 
our method have achieved  comprehensive statistical inference for the 11-dimensional parameters of MBHBs. 
This approach, as opposed to previous AI-based inference methods, is featured by  faster training and more accurate estimation.  

Besides, we propose the integration of model generation strategy based on the symmetry of the system with CNFs. 
This enables us to simplify the training task, while still obtaining a model that is applicable to the entire mission period.
Note that the coordinate transformation and CNFs are theoretically applicable to other space-based missions such as LISA.

It is also noteworthy that our method exhibits robustness against challenges such as confusion noise and the complexities of time-varying response function. The significant speed of our method, surpassing traditional techniques by orders of magnitude, positions it as a potential tool for the preprocessing of global fitting. Moreover, the precise and reliable estimation of sky location parameters (longitude, latitude) might be beneficial in guiding the search for electromagnetic observatories.

Overall, the acceleration in data processing and analysis not only streamlines workflows but also broadens the horizons for gravitational wave data exploration. These advancements herald a transformative era in space-based GW detection, fortifying the capabilities of missions like Taiji and bringing us closer to unraveling the mysteries of the universe.

Furthermore, 
the application of CNFs to other signals remains to be explored.
In this paper, CNFs are successfully applied to short-duration MBHB merger signals. However, the following two factors may pose challenges for the further applications of CNFs on other signals. Firstly, EMRIs and GBs typically require months of observation time to achieve considerable SNRs, hence significantly increasing the dimensionality of data. Secondly, EMRIs are known for their complicated waveforms and multiple local maxima on the likelihood surface~\cite{burke2024accuracyrequirementsassessingimportance}.

\section{Data access statement}
The data that support the findings of this study are available from the
corresponding author upon reasonable request. 

\section{Acknowledgements}
This study is supported by the National Key Research and Development Program of China (Grant No. 2021YFC2201901, Grant No. 2021YFC2203004, Grant No. 2020YFC2200100 and Grant No. 2021YFC2201903). International Partnership Program of the Chinese Academy of Sciences, Grant No. 025GJHZ2023106GC.

\bibliographystyle{unsrt}
\bibliography{main}

\end{document}